\documentclass[twocolumn,showpacs,preprintnumbers,prb,superscriptaddress]{revtex4}

\usepackage{graphicx}
\usepackage{dcolumn}
\usepackage{bm}
\usepackage{psfig}

\begin{document}

\title{Method for direct observation of coherent quantum oscillations in a
superconducting phase qubit}
\author{Ya. S. Greenberg}
\altaffiliation[On leave from ]{Novosibirsk State Technical
University, 20 K. Marx Ave., 630092 Novosibirsk, Russia.}
\author{A. Izmalkov}
\affiliation{%
Institute for Physical High Technology, P.O. Box 100239, D-07702
Jena, Germany\\
}%
\author{M. Grajcar}%
\altaffiliation[On leave from ]{Department of Solid State Physics,
Comenius University, SK-84248 Bratislava, Slovakia.}
\affiliation{Friedrich Schiller University, Department of Solid
State Physics, D-07743 Jena, Germany}
\author{E.  Il'ichev}
\email{ilichev@ipht-jena.de}
\affiliation{%
Institute for Physical High Technology, P.O. Box 100239, D-07702
Jena, Germany\\
}%
\author{W. Krech}
\affiliation{Friedrich Schiller University, Department of Solid
State Physics, D-07743 Jena, Germany}
\author{H.-G. Meyer}%
\affiliation{%
Institute for Physical High Technology, P.O. Box 100239, D-07702
Jena, Germany\\
}%

\date{\today}

\begin{abstract}Time-domain observations of coherent oscillations between
quantum states in mesoscopic superconducting systems were so far
restricted to restoring the time-dependent probability
distribution from the readout statistics. We propose a new method
for \emph{direct} observation of Rabi oscillations in a phase
qubit. The external source, typically in GHz range, induces
transitions between the qubit levels. The resulting Rabi
oscillations of supercurrent in the qubit loop induce the voltage
oscillations across the coil of a high quality resonant tank
circuit, inductively coupled to the phase qubit.  It is the
presence of these voltage oscillations in the detected signal
which reveals the existence of Rabi oscillations in the qubit.
Detailed calculation for zero and non-zero temperature are made
for the case of persistent current qubit. According to the
estimates for decoherence and relaxation times, the effect can be
detected using conventional rf circuitry, with Rabi frequency in
MHz range.
\end{abstract}

\maketitle


\section{Introduction}
 As is known the persistent current qubit (phase qubit) is one of
the candidates as a key element of a scalable solid state quantum
processor.\cite{Mooij1,Orlando} The basic dynamic manifestations
of a quantum nature of the qubit are macroscopic quantum coherent
(MQC) oscillations (Rabi oscillations) between its two basis
states, which are differed by the direction of macroscopic current
in the qubit loop.

Up till now the Rabi oscillations in time domain
\cite{Nakamura,Esteve} or as a function of the perturbation power
\cite{Han,Martinis} have been detected indirectly through the
statistics of switching events (e.g. escapes into continuum). In
either case the probabilitiy $P(t)$, or $P(E)$, was to be obtained
and analyzed to detect the oscillations.

More attractive in the long run is a direct detection of MQC
oscillations through a weak continuous measurement of a classical
variable, which would implicitly incorporate the statistics of
quantum switching events, not destroying in the same time the
quantum coherence of the qubit.\cite{Averin,Korotkov,Korotkov1}

In this paper we describe the approach which allows a direct
measuring of MQC oscillations of macroscopic current flowing in a
loop of a phase qubit. This qubit variety has the advantage of
larger tolerance to external noise, especially to dangerous random
background charge fluctuations.\cite{Blatter} To be specific, we
will use the example of three-junction small-inductance phase
qubit (persistent current qubit\cite {Orlando}) where level
anticrossing was already observed.\cite{Mooij}

\begin{figure}[tbp]
\includegraphics[width=7cm]{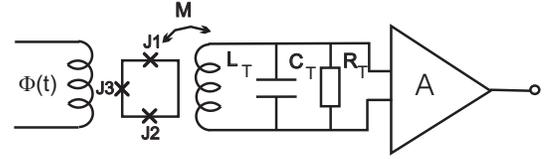}
\caption{Phase qubit coupled to a tank circuit.} \label{fig1}
\end{figure}

In our method a resonant tank circuit with known inductance
$L_{T}$, capacitance $C_{T}$ and quality factor $Q_T$ is coupled
with a target Josephson circuit through the mutual inductance $M$
(Fig.~\ref{fig1}). The method was successfully applied to a
three-junction qubit in classical regime,\cite {APL7} when the
hysteretic dependence of ground-state energy on the external
magnetic flux was reconstructed in accordance to the predictions
of Ref.~\onlinecite{Orlando}.

The phase qubit is biased by external magnetic flux $\Phi(t)$.
Assuming small qubit self-inductance, we neglect the shielding
current. Therefore the flux through the qubit loop is
\begin{equation}
\Phi(t) =\Phi _{x}+\Phi _{ac}(t),  \label{Phi}
\end{equation}
where $\Phi _{x}$ is a time independent external flux,
$\Phi_{ac}(t)$ is a monochromatic high frequency flux from the
external source.

If time-dependent external flux is applied to the qubit, the
latter will be in a time-dependent superposition of states
$|0\rangle$ and $|1\rangle$. If the frequency of external flux is
in resonance with the interlevel spacing of the qubit, the average
current in the qubit loop will oscillate with the frequency
$\Omega$ which depends on the amplitude of external flux and can
be made much smaller than the frequency of external signal. These
smaller oscillations which are called Rabi oscillations can be
detected with the aid of a high quality tank circuit coupled
inductively to the qubit loop.

Below we find the expression for the current in a qubit loop in
presence of electromagnetic resonant excitation for zero and non
zero temperatures. We show that for typical parameters of
three-junction phase qubit the current oscillations in a qubit
loop give rise to the voltage oscillations across the tank with
the amplitude which can be at the $\mu$V level that makes a direct
detection of Rabi oscillation possible.

\section{Quantum dynamics of 3JJ phase qubit}

Quantum dynamics of three-junction phase qubit has been studied in
detail in Ref.~\onlinecite{Orlando}. The qubit consists of a loop
with three Josephson junctions.  The loop has very small
inductance, typically in pH range. It insures effective decoupling
of qubit from external environment. Two Josephson junctions have
equal critical current $I_C$ and capacitance $C$, while the
critical current and capacitance of a third junction is a little
bit smaller, $\alpha I_C$, $\alpha C$ where $0.5<\alpha<1$. If the
Josephson coupling energy $E_J=I_C\Phi_0/2\pi$, where
$\Phi_0=h/2e$ is a flux quantum, is much more than the Coulomb
energy $E_C=e^2/2C$, then the phase of a Cooper pair wave function
is well defined. As was shown in Refs.~\onlinecite{Orlando,Mooij1}
in the vicinity of $\Phi=\Phi_0/2$ this system has two quantum
stable states which macroscopically differ by the direction of the
current circulating in a qubit loop. In the absence of high
frequency excitation the quantum properties of the qubit are
described by Hamiltonian (Eq. (12) in Ref.~\onlinecite{Orlando}):
\begin{equation}  \label{H_0}
H_0=\frac{P_\varphi^2}{2M_\varphi}+\frac{P_\theta^2}{2M_\theta}%
+U(f,\phi,\theta)
\end{equation}
where $P_\varphi=-i\hbar\partial/\partial\varphi$,
$P_\theta=-i\hbar\partial/\partial\theta$,
$M_\varphi=(\Phi_0/2\pi)^22C$,
$M_\theta=(\Phi_0/2\pi)^22C(1+2\alpha)$;
\begin{eqnarray}  \label{U}
U(f,\varphi,\theta)=E_J \left\{ 2+\alpha-2\cos\varphi\cos\theta \right. \nonumber \\
\left. -\alpha\cos\left[2\pi\left(
f+\frac{1}{2}\right)+2\theta\right]\right\}
\end{eqnarray}
Here $\varphi=(\varphi_1+\varphi_2)/2$,
$\theta=(\varphi_1-\varphi_2)/2)$, where $\varphi_1$, $\varphi_2$
are gauge-invariant phases of two Josephson junctions with equal
critical currents. In contrast to Ref.~\onlinecite{Orlando},
in~(\ref{U}) we define the flux bias $f=\Phi/\Phi_0-\frac{1}{2}$
as a small parameter measuring the departure from degeneracy. At
the degeneracy point $f=0$ potential energy ~(\ref{U}) shows two
minima with equal energies $\varepsilon_0$ at the points
$\varphi=0$, $\theta=\pm\theta_c,$ where $\cos\theta_c=1/2\alpha$.
The tunneling between the minima lifts the degeneracy leading to
the energy levels $E_\pm=\varepsilon_0\pm\Delta$ where $\Delta$ is
a tunneling matrix element between two minima. However, at the
degeneracy point the current in a qubit loop vanishes, so that it
is necessary to move a little bit away from this point. In order
to find the levels in the close vicinity of degeneracy point we
expand potential energy ~(\ref{U}) near its minima taking account
for linear terms in $f$ and quadratic terms in quantum variables
$\phi$, $\theta$. With the use of the technique described in
Ref.~\onlinecite{Orlando} we find the following expression for the
energies of two levels:
\begin{equation}  \label{E_pm}
E_\pm=\varepsilon_0\pm\sqrt{E_J^2f^2\lambda^2(\alpha)+\Delta^2}
\end{equation}
where we take offset at the degeneracy point, i. e., $\emph{f=0}$
corresponds to $\Phi_X=\Phi_0/2$;
\begin{equation}  \label{e_o}
\varepsilon_0=E_J\left(\left(2-\frac{1}{2\alpha}\right) +\sqrt{\frac{1}{%
\alpha(E_J/E_C)}}\left(1+\sqrt{2\alpha-1}\right)\right)
\end{equation}

\begin{eqnarray}  \label{lambda}
\lambda(\alpha)=&&\frac{\pi}{\alpha}\left(-\sqrt{4\alpha^2-1}+\sqrt{\frac{%
\alpha}{(E_J/E_C)}}\right.  \nonumber \\
&&\times\left.\left(\frac{2\alpha^2-1}{\sqrt{4\alpha^2-1}} +\frac{2\alpha^2+1%
}{\sqrt{2\alpha+1}(4\alpha^2-1)}\right)\right)
\end{eqnarray}
Expression~(\ref{E_pm}) differs from corresponding equation in
Ref.~\onlinecite{Mooij} by a factor $\lambda(\alpha)$ which
explicitly accounts for the dependence of the energies $E_\pm$ on
$\alpha$ and $E_J/E_C$. The stationary state wave functions
$\Psi_\pm$ are eigenfunctions of Hamiltonian $H_0$ :
$H_0\Psi_\pm=E_\pm\Psi_\pm$. They can be written as the
superpositions of the wave functions in the flux basis, $\Psi_L,
\Psi_R$ where $L$, $R$ stand for the left, right well,
respectively: $\Psi_\pm=\emph{a}_\pm\Psi_L+\emph{b}_\pm\Psi_R$.
\begin{equation}  \label{a_pm}
a_\pm=\frac{\Delta}{\sqrt{(\varepsilon_+-E_\pm)^2+\Delta^2}};b_\pm=\frac{%
\varepsilon_+-E_\pm}{\sqrt{(\varepsilon_+-E_\pm)^2+\Delta^2}};
\end{equation}
where $\varepsilon_+=\langle\Psi_L|H_0|\Psi_L\rangle$; $\varepsilon_-=%
\langle\Psi_R|H_0|\Psi_R\rangle$. For stationary states the current
circulating in a qubit loop can be calculated either as the average of a
current operator $\widehat{I_q}=I_C\sin(\varphi+\theta)$ over stationary
wave functions or as a derivative of the energy over the external flux:
\begin{equation}  \label{I_q}
I_q=\langle\Psi_\pm|\widehat{I_q}|\Psi_\pm\rangle=\frac{\partial
E_\pm}{\partial\Phi}=\pm
I_Cf\frac{\lambda^2(\alpha)}{\pi}\frac{E_J}{\hbar\omega_0}
\end{equation}
where $\hbar\omega_0=E_+-E_-$.

Suppose we apply to the qubit the excitation on a frequency close
to a gap frequency $\omega=\hbar\omega_0\approx 1$~GHz. The
corresponding perturbation term is then added to Hamiltonian
$H_0$: $H_{int}=V(\phi,\theta)\cos(\omega t)$ where
$V(\phi,\theta)$ in the vicinity of the left (right) minimum is as
follows:
\begin{eqnarray}  \label{V_pm}
V_{L,R}(\theta,\varphi)=&&E_J\frac{\pi}{\alpha}f_{ac}\left(\mp\sqrt{4\alpha^2-1}
\pm\varphi^2\frac{2\alpha^2-1}{\sqrt{4\alpha^2-1}}\right.  \nonumber \\
&&\left.\pm(\theta-\theta_C^\pm)^2\frac{2\alpha^2+1}{\sqrt{4\alpha^2-1}}
\right)
\end{eqnarray}
\begin{equation}  \label{theta_c}
\theta_C^\pm=\pm\theta_C+2\pi f\frac{1-2\alpha^2}{4\alpha^2-1}
\end{equation}
Throughout the paper we assume $f\geq 0$, so that the bottom of
the left well is higher than the bottom of the right well.
Accordingly, in Eqs.~(\ref{V_pm}) and (\ref{theta_c}) the upper
sign refers to right minimum, the lower sign refers to left
minimum. The quantity $f_{ac}$ in Eq.~(\ref{V_pm}) is the
amplitude of excitation field in flux units:
$f_{ac}=\Phi_{ac}/\Phi_0$. The high frequency excitation induces
the transitions between two levels which result in a superposition
state for the wave function of the system:
$\Psi(t)=C_+(t)\Psi_++C_-(t)\Psi_-$. The coefficients $C_\pm(t)$
are obtained from the solution of time dependent Schrodinger
equation with proper initial conditions for $C_\pm(t)$. We assume
that before the excitation the system was at the lower energy
level: $C_-(t=0)=1; C_+(t=0)=0$. The corresponding solution for
$C_\pm(t)$ in the rotating wave approximation is as follows:
\cite{Landau}
\begin{eqnarray}  \label{C}
C_-(t)&=&e^{-i\frac{E_-}{\hbar}t}e^{i\frac{\omega-\omega_0}{2}t}\left[
\cos\Omega t-i\frac{\omega-\omega_0}{2\Omega}\sin\Omega t\right] \nonumber \\
C_+(t)&=&e^{-i\frac{E_+}{\hbar}t}e^{-i\frac{\omega-\omega_0}{2}t}\left(-i
\frac{\Omega_r}{\Omega}\right)\sin\Omega t
\end{eqnarray}
where $\Omega_r=\frac{|\langle\Psi_+|V|\Psi_-\rangle}{2\hbar}$,
$\Omega=\sqrt{\frac{(\omega-\omega_0)^2}{4}+\Omega_r^2}$. At
resonance $(\omega=\omega_0)$ we get $\Omega=\Omega_r$. Taking
$\Psi_L$, $\Psi_R$ as ground state oscillatory wave functions in
the left, right well, respectively, we calculate matrix element
$\langle\Psi_+|V|\Psi_-\rangle$ and obtain for Rabi frequency:
\begin{equation}  \label{Omega}
\Omega_r=\frac{E_J}{\hbar}f_{ac}|\lambda(\alpha)|\frac{\Delta}{\hbar\omega_0}
\end{equation}
As is seen from Eq.~(\ref{Omega}) by a proper choice of excitation
power ($f_{ac}$ in our case) the frequency $\Omega_r$ can be made
much lower the gap frequency $\Delta/h$. Now we calculate the
average current in a superposition state:
\begin{eqnarray}  \label{I_qpsi}
I_q&=&\langle\Psi(t)|\widehat{I_q}|\Psi(t)\rangle  \nonumber \\
&=&|C_+(t)|^2\langle\Psi_+|\widehat{I_q}|\Psi_+\rangle+|C_-(t)|^2\langle%
\Psi_-|\widehat{I_q}|\Psi_-\rangle \ \ \
\end{eqnarray}
where we neglect high frequency $\omega_0$ term. Accounting for
the expression (\ref{I_q}) for the current in stationary states we
obtain at resonance the following expression for the average
current flowing in a qubit loop:
\begin{equation}  \label{I_qI_c}
I_q=-I_Cf\frac{\lambda^2(\alpha)}{\pi}\frac{E_J}{\hbar\omega_0}\cos2\Omega_rt
\end{equation}

At finite temperature the calculation  of the current in a qubit
loop in presence of high frequency excitation is based on the
density matrix equation: $ \label{rhot}
i\hbar\dot{\rho}(t)=\left[\left(H_0+H_{int}(t)\right),\rho(t)\right]$
with initial conditions at thermal equilibrium:
$\rho_{++}(0)=\rho_{++}^{eq}$; $\rho_{--}(0)=\rho_{--}^{eq}$;
$\rho_{+-}(0)=\rho_{-+}(0)=0$, where $\rho_{++}^{eq}$,
$\rho_{--}^{eq}$ are equilibrium density matrix elements :
$\rho_{++}^{eq}=\frac{1}Z{e^{-\frac{E_+}{k_BT }}}$;
$\rho_{--}^{eq}=\frac{1}Z{e^{-\frac{E_-}{k_BT}}}$,
$Z=e^{-\frac{E_+}{k_BT}}+e^{-\frac{E_-}{k_BT}}$.

In the rotating wave approximation the diagonal elements of
density matrix at resonance ($\omega=\omega_0$) are as follows:
\begin{equation}  \label{rho_mm}
\rho_{--}(t)=\frac{1}{2}+\frac{1}{2}\tanh\frac{\hbar\omega_0}{2k_bT}\cos2\Omega_rt,
\end{equation}
$\rho_{++}(t)=1-\rho_{--}(t)$.

Now we find the average current at resonance for nonzero
temperature:
\begin{eqnarray}  \label{Iq_Ej}
I_q&=&\langle\Psi_+|\widehat{I_q}|\Psi_+\rangle\rho_{++}(t)+\langle\Psi_-|%
\widehat{I_q}|\Psi_-\rangle\rho_{--}(t)  \nonumber \\
&=&-I_C\frac{E_Jf\lambda^2(\alpha)}{\pi\hbar\omega_0}\tanh\left(\frac{%
\hbar\omega_0}{2k_BT}\right)\cos2\Omega_rt
\end{eqnarray}

In order to estimate the Rabi frequency $\Omega_r$  we take the
following qubit parameters: $I_C=400$~nA, $\Delta/h=0.3$~GHz,
$\alpha=0.8$, $L=15$~pH, $E_J/E_C=100$. We take the amplitude of
time-dependent flux which is coupled to qubit from high frequency
source $f_{ac}=1\times10^{-4}$. We set the flux offset from
degeneracy point $f=3.5\times10^{-4}$, so that
$\hbar\omega_0=2\sqrt{2}\Delta$, $\omega_0/2\pi=0.85$~GHz. For
these values we obtain from Eq.~(\ref {Omega})the Rabi frequency
$\Omega_r/2\pi=32$~MHz.

\section{The interaction of the phase qubit with a tank circuit}

The problem of a coupling a quantum object to the classical one,
which is dissipative in its nature, has no unique theoretical
solution. A rigorous approach is to start from exact Hamiltonian
which describes the qubit-tank circuit system:
\begin{equation}\label{H}
H=H_0+H_T+H_{0T}+H_{TB}+H_B
\end{equation}
where $H_0$ is the qubit Hamiltonian given in Eq.~(\ref{H_0}),
$H_T$ is the tank circuit Hamiltonian
\begin{equation}\label{H_T}
  H_T=\frac{Q^2}{2C_T}+\frac{\Phi^2}{2L_T}
\end{equation}
where $Q$ and $\Phi$ are a quantum operators of the charge at the
capacitor and of magnetic flux trapped by the inductor of a tank
circuit, respectively. The operators obey commutator relation
$[\Phi,Q]=i\hbar$.  The interaction Hamiltonian between the qubit
and the tank, $H_{0T}$ is:
\begin{equation}\label{H_{0T}}
  H_{0T}=\frac{M}{L}\hat{I}_q\Phi
\end{equation}
where $M$ is inductive coupling between qubit and the tank; $H_B$
is the Hamiltonian of a thermal bath coupled to the tank via
interaction $H_{TB}=\alpha\Phi\Gamma$, where $\alpha$ is the
coupling constant between the tank and dissipative environment,
$\Gamma$ is the dynamic variable of thermal bath $H_B$.

The equations of motion for tank circuit variables are as follows:
\begin{equation}\label{dQ/dt}
  \frac{dQ}{dt}=-\frac{\Phi}{L_T}-\frac{M}{L}\hat{I}_q+\alpha\Gamma
\end{equation}
\begin{equation}
  \frac{d\Phi}{dt}=\frac{Q}{C_T}
  \label{dPhi/dt}
\end{equation}
From these two equations we get for the voltage operator
$\hat{V}=Q/C_T$ across the tank:
\begin{equation}\label{Volt}
\frac{d^2\hat{V}}{dt^2}+\omega_T^2\hat{V}=
-M\omega_T^2\frac{d\hat{I}_q}{dt}+\alpha\frac{d\Gamma}{dt}
\end{equation}
The averaging of this equation over the bath leads to the
dissipative equation for the average voltage, $V$, across the tank
(see, for example, Ref.~\onlinecite{Weiss}):
\begin{equation}\label{avVolt}
\ddot{V}+\frac{\omega_{T}}{Q_T}\dot{V}+\omega_{T}^{2}V=
-M\omega_{T}^{2}\frac{dI_{q}}{dt}.
\label{main1}
\end{equation}
where $Q_T=\omega _{T}R_{T}C_{T}\gg 1$ is the tank quality factor,
$\omega _{T}=1/\sqrt{L_{T}C_{T}}$.

In the spirit of selective quantum evolution
approach,\cite{Korotkov} we interpret Eq.~\ref{avVolt} as follows.
Suppose the qubit is in a pure state, $a\left| 0\right\rangle
+b\left| 1\right\rangle .$ The tank voltage is measured (in
quantum-mechanical sense) at certain times $t_{k}=\Delta t,2\Delta
t,...$. Each time the qubit state is also measured, since there is
correspondence between $V(t_{k})$ and qubit current, which thereby
takes value either
$\left\langle0\right|\hat{I}_{q}(t_{k})\left|0\right\rangle$ or
$\left\langle 1\right|\hat{I}_{q}(t_{k})\left| 1\right\rangle$,
with appropriate probability. Averaging $V(t_{k})$ over intervals
$\Delta T\gg \Delta t,$which are still small compared to other
characteristic times in the system, would yield Eq.~\ref{avVolt}.

The quantity to be detected is the oscillating voltage across the
tank which at resonance is $V=V_a\cos2\Omega_rt$, where the
voltage amplitude, $V_a$ is:
\begin{equation}  \label{V}
V_a=MQI_Cf\frac{2\lambda^2(\alpha)}{\pi}\frac{E_J}{\hbar\omega_0}%
\Omega_r.
\end{equation}
 It is the presence of these voltage
oscillations in the detected signal which reveals the existence of
Rabi oscillations in the qubit.

The Eq.~(\ref{avVolt}) is still quantum equation since the average
voltage $V$ is a quantum operator of the voltage across the tank
averaged over the bath degrees of freedom.  Therefore, our problem
is reduced to the problem of measuring a weak external force (in
our case, $MdI_q/dt$) by a dissipative oscillator. As is known
(see, for example, Ref.~\onlinecite{Braginsky}) a classical
descriptions of such oscillator requires that the quantum
fluctuations of the detector variable (i. e. $V$ in our case) in
the measurement bandwidth
 be smaller than the amplitude induced in the tank coil by external signal,
$MQ_TdI_q/dt$. According to the fluctuation-dissipation theorem
the quantum fluctuations of the voltage $V$ are given by the
spectral density:
\begin{equation}\label{spd}
    S_V(\omega)=2ReZ(\omega)\hbar\omega\coth\left(\frac{\hbar\omega}{2k_BT}\right)
\end{equation}
where $Z^{-1}(\omega)=(1/R_T+1/i\omega L_T+i\omega C_T)$ is the
impedance of a tank circuit.

Below we take the following parameters of the tank circuit:
$L_T=50$~nH, $Q=1000$, inductive coupling to qubit,
$k^2=M^2/LL_T=10^{-4}$. We assume the tank is tuned to the
frequency of oscillating current in the qubit, i. e.,
$\omega_T/2\pi=2\Omega_r/2\pi=64$~MHz, hence, $C_T=124$~pF. We
take the flux amplitude which is coupled to qubit from high
frequency source $f_{ac}=1\times10^{-4}$. We set the flux offset
from degeneracy point $f=3.5\times10^{-4}$, so that $\hbar\omega_0=2\sqrt{2}%
\Delta$, $\omega_0/2\pi=0.85$~GHz. For these values we obtain for
the voltage amplitude at resonance $V_a\approx$~0.7~$\mu$V.

From Eq.~(\ref{spd}) we estimate the voltage fluctuations across
the tank circuit coil at resonance ($\omega=\omega_T$) :
\begin{equation}  \label{V_n}
V_n=\sqrt{\frac{S_V(\omega_T)}{Z(\omega_T)}}\omega_TL_TQ_T\sqrt{B}
\end{equation}

where $B=\omega_T/2\pi Q_T$ is the bandwidth of the tank circuit.
We perform the calculations for $T=10$~mK. For the voltage
fluctuations we get $V_n\approx 10$~nV. Thus, we see that the
voltage fluctuations across the tank coil is much smaller than the
signal amplitude from the qubit. Therefore, we may treat the tank
circuit as a classical object and the voltage $V(t)$ as the
classical variable coupled through Eq.~\ref{avVolt} to the qubit
degree of freedom, where the current $I_{q}$ in Eq.~\ref{avVolt}
is calculated as the average of a current quantum operator over
the qubit statistical operator.

\section{The effects of qubit relaxation and decoherence}
 The estimations we made above are very promising, however,
our derivation is made under a strong assumption: we neglected the
decoherence due to interaction of the qubit with external
environment and with a  measuring device. In fact, the possibility
of detection of Rabi frequency depends crucially on relaxation,
$\Gamma_r$, and decoherence, $\Gamma_\varphi$ rates, which lead to
the decay of Rabi oscillations. The detection is in principle
possible if period of oscillations is small compared to
$min[1/\Gamma_\varphi, 1/\Gamma_r]$. The decoherence is caused
primarily by coupling of a solid state based phase qubit to
microscopic degrees of freedom in the solid. Fortunately this
intrinsic decoherence has been found to be quite
weak:\cite{LinTian} the intrinsic decoherence times appeared to be
on the order of 1 ms which is several orders of magnitude more
than period of current oscillations we estimated before:
$\pi/\Omega_r=15.6$~ns. However, the external sources of
decoherence are more serious. In our method these are the
microwave source which induces the Rabi oscillations and the tank
circuit which has to detect them. From this point two structures
of microwave source has been analyzed: coaxial line that is
inductively coupled to the qubit\cite{Wal} and on-chip oscillator
based on overdamped DC SQUID.\cite{Crankshaw} The analysis has
shown the relaxation and decoherence rates were on the order of
100~$\mu$s at 30~mK for coaxial line, and 150~$\mu$s and
300~$\mu$s for relaxation and decoherence, respectively, for
on-chip oscillator at 1~GHz. Here we estimate decoherence times
that are due to a tank circuit using the expressions for
$\Gamma_r$ and $\Gamma_\varphi$ from\cite{Grifoni}
\begin{equation}\label{Gamma_r}
\Gamma_r\equiv\frac{1}{T_r}=\frac{1}{2}\left(\frac{\Delta}{\hbar\omega_0}%
\right)^2J(\omega_0) \coth\left(\frac{\hbar\omega_0}{2k_BT}\right)
\end{equation}
\begin{equation}\label{Gamma_phi}
\Gamma_\varphi\equiv\frac{1}{T_\varphi}=\frac{\Gamma_r}{2}+2\pi\eta\left(%
\frac{E_Jf\lambda}{\hbar\omega_0}\right)^2 \frac{k_BT}{\hbar}
\end{equation}
where dimensionless parameter $\eta$ reflects the Ohmic
dissipation. It depends on the strength of noise coupling to the
qubit. In our subsequent estimations we take $\eta=5\times
10^{-3}$ which is relevant for weak damping limit. Below we use
the approach described in Ref.~\onlinecite{LinTian1} in its
simplified form.\cite{Wal,Crankshaw,Orlando1} While it is not
mathematically rigorous, nevertheless it gives a correct order of
magnitude of relaxation times. The quantity $J(\omega_0)$ is the
zero temperature spectral density of the fluctuations of the gap
energy of the qubit:
$J(\omega)=<\delta\varepsilon(\omega)\delta\varepsilon(\omega)>/\hbar^2$.
The fluctuations $\delta\varepsilon$ are due to the flux noise
$\delta f$, which is supplied to the qubit by a tank circuit. From
Eq.(5) we get
$\delta\varepsilon=\left(4E_J^2\lambda^2f/\hbar\omega_0\right)\delta
f$, where $\delta f=M\delta I/\Phi_0$. The current noise $\delta
I$ in a tank circuit inductance comes from two independent parts:
Johnson-Nyquist voltage noise in a tank circuit resistance with a
spectral density at $T=0$: $S_V(\omega)=2\hbar\omega ReZ(\omega)$,
where $Z^{-1}(\omega)=(1/R_T+1/i\omega L_T+i\omega C_T)$ is the
impedance of a tank circuit, and from a current noise of
preamplifier with a spectral density $S_A$. Therefore, we get for
$J(\omega_0)$:
$J(\omega_0)=\left(4E_J^2\lambda^2fM/\hbar^2\omega_0\Phi_0
\right)^2S_I(\omega_0)$, where $S_I(\omega)=\left[S_V(\omega)ReZ(\omega)+S_A%
\left|Z(\omega)\right|^2\right]/\omega^2L^2$ is the spectral
density of a current noise in the tank circuit inductance. Since
$\omega_0>>\omega_T,
S_I(\omega_0)\cong\left(2\hbar\omega_0/R_T+S_A\right)\left(\omega_T/\omega_0%
\right)^4$. For the estimation we take $T=10$~mK,
$R_T=20$~k$\Omega$, $S_A=10^{-26}$~A$^2$/Hz,
$\omega_T/2\pi=64$~MHz, other parameters being the same as before.
We find that the contribution to the relaxation of a tank circuit
noise and of preamplifier noise is approximately 0.1~s$^{-1}$ and
22~s$^{-1}$, respectively. Therefore total relaxation is
determined by preamplifier noise giving relaxation time
$T_r\cong45$~ms. The decoherence rate is dominated by a second
term in (\ref{Gamma_phi}), which is equal $5.4\times
10^6$~s$^{-1}$ giving decoherence time $T_\varphi=185$~ns, which
is approximately ten oscillation periods of circulating current.
These estimations clearly show the possibility of detection of
Rabi frequency in MHz range with the aid of conventional rf
circuitry.

Here we did not consider the effect of a tank circuit back action
on the qubit which can give additional contribution to the
relaxation and decoherence rates. In order to reduce the back
action effect of a tank circuit  and to further enhance the MQC
signal it may be advantageous for detection of Rabi oscillations
to use a two dimensional array of identical phase qubits coupled
to a tank circuit. Modern technology allows several thousands of
weakly coupled phase qubits to be obtained on a chip.\cite{Wal}
Preliminary experiments showed that the behavior of macroscopic
current in such a system is completely analogous to that of
longitudinal magnetization in NMR: the collective reversal of the
persistent currents in the qubit loops when sweeping the flux bias
within a degeneracy point has been observed.\cite{Wal}

In conclusion, we have shown that Rabi oscillations (in MHz range)
of the current circulating in a qubit loop which are induced by
high frequency external source (in GHz range) can be detected in
MHz range as the voltage oscillations in the high quality tuned
tank circuit inductively coupled to the qubit.

The authors thank D-wave Sys. Inc. for partial support and A.
Zagoskin, A. M. van den Brink, M. Amin, A. Smirnov, N. Oukhanski,
and M. Fistul for useful discussions.

\end{document}